\documentstyle[11pt]{article}
\title{A DLA model for Turbulence}
\author{Asher Yahalom$^{a}$  \\
$^a$Ariel University, Ariel 40700, Israel\\
e-mail:  asya@ariel.ac.il; }

\newcommand{\beq} {\begin{equation}}
\newcommand{\enq} {\end{equation}}

\newcommand {\eqref}[1] {equation (\ref{#1})}
\newcommand {\ern}[1] {equation (\ref{#1})}

\begin{document}
\maketitle

\begin{abstract}
A connection between fractal dimensions of "turbulent facets" and
fractal dimensions in diffusion-limited aggregation (DLA) is
shown. The theoretical correspondence is elucidated and an
empirical support to the above claim is given.

\end{abstract}

\section{Introduction}

In many turbulent shear flows such as boundary layers, jets, mixing layers and wakes there is
a sharp interface that divides the flow field into two distinct regions. In one region the
flow is turbulent while in the other, the flow is largely a potential flow (\cite{CoKi,Silva1}). This sharp
interface - the turbulent/nonturbulent interface  is continually deformed over a wide
range of scales and the flow dynamics in its vicinity determines many of the most important
flow features: the growth and spreading rate of wakes, the exchanges of mass across mixing
layers, and the mixing and reaction rates in jets are some of the flow features that are largely
determined by the characteristics of the interface and the flow dynamics in its vicinity.
The key event that occurs at a is the `communication' of vorticity from the core of the
turbulent region into the irrotational zone. Turbulent entrainment can be seen as the mechanism
by which fluid elements from the irrotational flow region acquire vorticity and become part of
the turbulent region. Past studies described the entrainment as being caused by large-scale eddy
motions (engulfment) occurring from time to time at particular locations along the interface \cite{Townsend}, but
recent works suggest instead that the entrainment results from small scale motions (nibbling)
acting along the entire interface \cite{Mathew,Hunt}, as originally described by \cite{CoKi}.
This process can also be viewed as a motion of the interface between potential and turbulent flow, which
will be the approach followed in this paper.

    In a number of articles (see e.g. \cite{man1}, \cite{man2}, \cite{ser} and later \cite{Silva2})
 the fractal dimensions of "turbulent facets" were considered. Among them is the fractal dimension of
 the interface surface between the potential flow region and the turbulent flow region \cite{ser}.
Such an interface exists in most classical turbulent flows such as axisymmetric jets, plane wakes and
 boundary layer turbulence. The fractal dimensions Sreenivasan et al. \cite{ser} measured are between
 2.3-2.4 with an error of 0.04 (see table-1 for results in 1-D and 2-D slicing\footnote{Cutting a d-m dimensional slice (cross section)
  of a D dimensional fractal embedded in a d-dimensional space leads to a D-m dimensional object.
Example: A 1-dimensional cut through the 1.46 dimensional checker board fractal is expected to have dimension 0.46. This means
that using a slice one is able to estimate the original fractal dimension of the sliced object.}).
 Dimension was measured in the interval between the Kolmogorov scale up to approximately the scale of
 large eddies (above this scale fractal dimension was found to be not well defined).
 It is claimed that the dimension of those interfaces are close and in some flows similar
(within the error limit) to the dimension of the interface generated in the DLA model.

\section{DLA fluid correspondence}

It is well-known that the problem of diffusion-limited aggregation
(DLA) is useful for describing various  phenomena such as viscous
fingering \cite{Pa}, electrical discharge \cite {Ni}
 and other examples \cite{Fal}.
This model provides a solution to the Laplace equation \cite{Ta} for the function $u$ whose physical meaning depends on
the phenomena under study:
\begin{eqnarray}
\triangle u = 0
\label{eq1}
\end{eqnarray}
under the following boundary conditions: $u=const$ on some far "rigid" boundary and a normal velocity $v_{n}$
on a moving boundary F, which moves with the velocity $v$
proportional to the derivative of $u$ in the direction $\hat n$ normal to F plus a random perturbation $p$.
 Thus, for a constant $k$ (which depend on the specific problem in hand, but does not alter
 the results that concerns us):
\begin{eqnarray}
 v_{n} = k \hat n \cdot \vec \nabla u + p
\label{eq2}
\end{eqnarray}
It is shown in the appendix  that the above boundary conditions determine $u$ uniquely, as the normal derivative of 
$u$ is determined on the surface by: $\hat n \cdot \vec \nabla u = \frac{v_{n}-p}{k}$.
The surface  F becomes a thin fractal under the above equations
with dimension of 2.41 in the 3-dimensional case measured by 2-D slicing \cite{Me}.

    Dimensional similarity does not indicate what kind of
 mechanism or equations generated this type of surface, however, the equations of potential flow
 outside the turbulent region and their boundary conditions are
quite the same as the ones the DLA model should solve. In this region the velocity field is given by:
\begin{eqnarray}
 \vec v = \vec \nabla u
\label{eq3}
\end{eqnarray}
where $u$ is the velocity potential. Since in an incompressible liquid the velocity field must obey the continuity equation of the form:
\begin{eqnarray}
\vec \nabla \cdot \vec v = 0
\label{eq4}
\end{eqnarray}
we obtain \ern{eq1}. The normal component of the velocity field must certainly be zero on the
 surface of the vessel containing the liquid, thus on this surface $u$ must be constant satisfying
the first boundary condition. The boundary surface between turbulent and potential flow moves very
 rapidly and erratically. This follows from the fact that the normal velocity
  to the surface {\bf is} the velocity of the surface. Thus satisfying the second boundary
condition. The velocity of the boundary is of the form \ern{eq2} with
$k=1$ and where $p$ is due to random deviation from pure potential flow.
The deviations of the flow from pure potential flow can be treated
as a deterministic chaos effect but it also can be taken as a
random perturbation due to complexity and smallness. We exclude
the influence of large eddies since our description is implied
only  up to the scale of this eddies.

\section{Empirical results}

Based on previous section reasoning we may conclude that the DLA
model describes the potential flow near a turbulent boundary,
since it  satisfies the same equations with the same boundary
conditions. Hence we are led to believe that the fractal dimension
of the interface surface generated by the DLA model  (2.41)
\cite{Me} should be in agreement with the fractal dimension of
experimental turbulent interfaces as given in table 1 below. This
is indeed the case.

\begin{bf}
Fractal dimension of interface
\end{bf}
\begin{tabbing}
Axisymmetric jet       \=2-D slicing        \=1-D slicing    \kill
Flow \                 \>2-D slicing        \>1-D slicing   \\
Boundary layer\        \>2.38               \>2.40          \\
Axisymmetric jet\      \>2.33               \>2.32           \\
Plane wake       \     \>----               \>2.37           \\
Mixing layer     \     \>----               \>2.40           \\
\end{tabbing}
Table 1. Summary of the fractal dimensions of the turbulent/ non-turbulent interface in several
 classical flows \cite{ser}.

\section{Kolmogorov's Model}

For turbulence, the physics appears to be that there is a scale-dependent eddy-diffusion that follows Kolmogorov famous scaling of $5/3$:
\beq
E(k) = C \varepsilon^{2/3} k^{-5/3}
\enq
$E(k)$ is the energy spectrum, $\varepsilon$ is the energy dissipation rate and $k$ is the wave number inversely proportional to the wave length.
This has much to do with the kinetic energy distribution among different scales, but no clear connection to the
 geometric dimension of the interface between were the eddies are created (boundary layer) and the potential flow zone.
 There may be a connection between the geometric dimension which is about $7/3$ and  Kolmogorov scaling of $5/3$.

\section{Conclusion}

We have shown that there is an excellent agreement between the
fractal dimension of the turbulent/ non-turbulent interface in
several classical flows and the fractal dimension of a DLA interface.

Notice that we do not intend to explain the phenomena of turbulence, only to offer a simple
 model for a specific facet of this complex phenomena, which apparently works very well.

\appendix

\section*{Appendix: Uniqueness of $u$}
\label{Uniqueness}

Let us consider a potential flow such that $\vec{v}=\vec \nabla u$. If the flow is incompressible we have: $\vec \nabla  \cdot \vec{v}=0$, hence: \beq
{\vec \nabla}^2 u=0
\enq
Now consider two possible solutions of the above equation, say ${ u}_1$ and ${ u}_2$, the difference between the solutions is also a solution: $\bar u ={ u}_1 - { u}_2$, hence we have:
\beq
{\vec \nabla}^2\bar{ u}=0
\enq
and also:
\beq
\bar{ u} {\vec \nabla}^2\bar{ u}=0
\enq
This means that:
\beq
\vec \nabla\cdot \left(\bar{ u}\vec \nabla \bar{ u}\right)-{\left(\vec \nabla \bar{ u}\right)}^2=0
\enq
Now take a volume integral of this:
\beq
 \int{\left[\vec \nabla\cdot \left(\bar{ u}\vec \nabla \bar{ u}\right)-{\left(\vec \nabla \bar{ u}\right)}^2\right]dx^3}=0
 \enq
And use Gauss theorem to obtain:
\beq
\oint{\bar{ u}\vec \nabla \bar{ u}}\cdot d\vec{S}=\int{{\left(\vec \nabla \bar{ u}\right)}^2 dx^3}
\enq
the left hand integral is over a surface encapsulating the volume of the right hand integral. This means that if either $\bar{ u}$ or the normal component $\vec \nabla \bar{ u}\cdot \hat{n}$ vanish on such a surface so must:
\beq
\int{{\left(\vec \nabla \bar{ u}\right)}^2dx^3=0}
\enq
And hence $\vec \nabla \bar{u}$ must be null throughout the volume. Which means that $\bar{u}$ is a spatial constant throughout the volume. This in turn means the if the potential of the flow is given on the encapsulating surface or the \textbf{normal component of the velocity} is given on such a surface than the flow is determined uniquely up to a spatial constant. This constant is determined uniquely if potential boundary conditions are given in at least one point of the surface.


\begin {thebibliography} {99}

\bibitem {CoKi} S. Corrsin  and A. L. Kistler,  1955, Free-stream Boundaries of Turbulent Flows Tech. Rep. TN-1244 NACA.
\bibitem {Silva1} C. B. da Silva, R. R. Taveira and G. Borrell, 2014,
Journal of Physics: Conference Series 506, 012015 doi:10.1088/1742-6596/506/1/012015
\bibitem {Townsend} A. A. Townsend, 1966, The mechanism of entrainment in free turbulent flows J. Fluid Mech. {\bf 26} 689-715.
\bibitem{Mathew}
J. Mathew  and A. Basu, 2002, Some characteristics of entrainment at a cylindrical turbulent boundary Phys. Fluids {\bf 14} 2065-72
\bibitem{Hunt}
J. Westerweel , C. Fukushima, J. M. Pedersen and J. C. R. Hunt, 2005, Mechanics of the turbulent-nonturbulent
interface of a jet Phys. Rev. Lett. {\bf 95} 174501
\bibitem {man1} B.\ B.\ Mandelbrot, 1974, J. Fluid Mech 62, 331.
\bibitem {man2} B.\ B.\ Mandelbrot, 1975, J. Fluid Mech 72, 401.
\bibitem {ser} K.\ R.\ Sreenivasan and C.\ Meneveau, J. Fluid Mech. 173, 357 (1986).
\bibitem {Silva2}
C. M. de Silva,  J. Philip, K. Chauhan , C. Meneveau  and I. Marusic,  2013, Multiscale geometry and scaling of the
turbulent-nonturbulent interface in high Reynolds number boundary layers Phys. Rev. Lett. 111 044501.
\bibitem {Fal} K.\ Falconer, 1990, Fractal Geometry p.267,  John Wiley \& Sons.
\bibitem {Me} P.\ Meakin Phys. Rev. A 27, 1495-1507. (1983)
\bibitem {Ni} L.\ Niemeyer L.\ Pietronera \& H.\ J.\ Wiesmas Phys. Rev. Lett. 52 (1984) 1033-1036.
\bibitem {Pa} L.\ Paterson Phys. Rev. Lett. 52 (1984)
\bibitem {Ta} T.\ A.\ Witten L.\ M.\ Sander Phys. Rev. B 27, 5686 (1983)

\end{thebibliography}
\end{document}